\documentclass{article}

\usepackage{arxiv}

\usepackage[utf8]{inputenc} % allow utf-8 input
\usepackage[T1]{fontenc}    % use 8-bit T1 fonts
\usepackage{hyperref}       % hyperlinks
\usepackage{url}            % simple URL typesetting
\usepackage{booktabs}       % professional-quality tables
\usepackage{amsfonts}       % blackboard math symbols
\usepackage{nicefrac}       % compact symbols for 1/2, etc.
\usepackage{microtype}      % microtypography
\usepackage{lipsum}
\usepackage{graphicx}
\graphicspath{ {./images/} }
\usepackage{algorithm2e}
\usepackage{appendix}
\usepackage{url}
\usepackage{subfigure}

\usepackage[
    style=numeric,
    sorting=none
]{biblatex}
\usepackage{algorithm2e}
\usepackage{listings}
\usepackage{graphicx}
\usepackage[T1]{fontenc}
\usepackage[scaled=0.85]{beramono}
\usepackage{listings}
\usepackage{multirow}
\usepackage{pgfplots}
\pgfplotsset{width=10cm,compat=1.9}

\usepgfplotslibrary{external}
\usepackage{svg}

\usepackage{pgfplots}
\usepgfplotslibrary{dateplot} 
\usetikzlibrary{calc}

\usepackage{algorithm2e}
\usepackage{listings}
\usepackage{graphicx}
\usepackage[T1]{fontenc}
\usepackage[scaled=0.85]{beramono}
\usepackage{listings}
\makeatletter
\long\def\ifnodedefined#1#2#3{%
    \@ifundefined{pgf@sh@ns@#1}{#3}{#2}%
}
\makeatother

\title{Linked Data on Geo-annotated Events and Use Cases for the Resilience of Ukraine}

\author{
Manar Attar\\
  Dept. of Computer Science\\
  Vrije Universiteit Amsterdam, \\Amsterdam, the Netherlands \\
\texttt{manarattar77@gmail.com}\\ 
   \And
 Shuai Wang\\
Dept. of Computer Science\\
  Vrije Universiteit Amsterdam, \\Amsterdam, the Netherlands \\
  \texttt{shuai.wang@vu.nl} \\
  \And
Ronald Siebes \\
Dept. of Computer Science\\
  Vrije Universiteit Amsterdam,\\ Amsterdam, the Netherlands \\
  \texttt{r.m.siebes@vu.nl} \\
  \And
Eirik Kultorp\\
Unaffiliated, Norway \\
  \texttt{eirik@ekul.no} 
  \And
Zhisheng Huang\\
Dept. of Computer Science\\
  Vrije Universiteit Amsterdam,\\ Amsterdam, the Netherlands \\
  \texttt{z.huang@vu.nl} \\
  \And
  Tianyang Lu\\
Unaffiliated, China \\
  \texttt{angelalu0818@gmail.com} 
}

\begin{document}
\maketitle
\begin{abstract}
The mission of resilience of Ukrainian cities calls for international collaboration with the scientific community to increase the quality of information by identifying and integrating information from various news and social media sources. Linked Data technology can be used to unify, enrich, and integrate data from multiple sources. In our work, we focus on datasets about damaging events in Ukraine due to Russia's invasion between February 2022 and the end of April 2023. We convert two selected datasets to Linked Data and enrich them with additional geospatial information. Following that, we present an algorithm for the detection of identical events from different datasets. Our pipeline makes it easy to convert and enrich datasets to integrated Linked Data. The resulting dataset consists of 10K reported events covering damage to hospitals, schools, roads, residential buildings, etc. Finally, we demonstrate in use cases how our dataset can be applied to different scenarios for resilience purposes.\footnote{The paper is an extended version of our 2023 paper titled `Converting and Enriching Geo-annotated Event Data: Integrating Information for Ukraine Resilience'\cite{sigspatial} presented at the ACM International Conference on Advances in Geographic Information Systems (SIGSPATIAL conference). It was included in the \textit{SIGSPATIAL '23: Proceedings of the 31st ACM International Conference on Advances in Geographic Information Systems}. The use cases were presented at the BNAIC-BeNeLearn Joint International Scientific Conferences on A.I. and Machine Learning. This long paper is the result of a merge of them with some minor extensions and additional explanations. Please cite our SIGSPATIAL paper instead. Its DOI is 10.1145/3589132.3625580. Related information and resources can be found on the Linked4Resilience platform (\url{https://www.linked4resilience.eu}).}

\end{abstract} 

% The integrated dataset remains available upon request while other intermediate datasets are available on the TriplyDB service with 
 % related resources available on the Linked4Resilience platform} (\url{https://www.linked4resilience.eu})

% keywords can be removed
\section{Introduction}
\label{section:introduction}
% general intro
The Russian-Ukrainian conflict since February 2022 has damaged civilian infrastructure, facilities, and buildings, sparking a large displacement crisis in Europe.
% The Russian-Ukrainian conflict escalated to war level in February 2022 when Russia launched a full-scale invasion of Ukraine. Enormous casualties and destruction on the country's infrastructure has take place by the constant bombardment of cities, damaging civilian infrastructure and buildings, and sparking a large displacement crisis in Europe. 
 It was reported in 2022 that, in Kharkiv alone, the war has resulted in large-scale destruction of infrastructure with an estimation of more than 1,000 buildings destroyed, among which 700 are multi-storey apartment buildings and are no longer habitable \cite{Chumachenkoo796}. Rebuilding destroyed residential buildings, public facilities, and social infrastructure is critical for the resilience of Ukraine. It is no easy task and will require international cooperation and coordination for reconstruction, including integration and management of datasets and resources of various kinds.
% decisions about resilience rely on datasets
 There are many open datasets reporting the progress of the Russian-Ukrainian conflict from various perspectives. These datasets enable researchers and analysts to gain insights into the complex situation, ultimately contributing to the development of effective strategies to protect civilians, promote peace, estimate resources for projects for resilience, etc.
% WikiEvent:
WikiEvents \cite{wikievents} consists of entries automatically curated based on Wikipedia’s Current Events portal.\footnote{\url{https://en.wikipedia.org/wiki/Portal:Current_events}} Its NLP downstream pipeline extracts 21,275 events including around a thousand events about the Russian invasion of Ukraine. ACLED is a much larger dataset\footnote{\url{https://acleddata.com/2023/03/01/war-in-ukraine-one-year-on-nowhere-safe/}, visited on 13th June, 2023.} with over one million events, including around 40,000 political violence events across Ukraine \cite{acled}. This dataset is mostly dedicated to military-related events about shelling, artillery, and missile strikes.
% public datasets for Russo-Ukraine conflict
 The Centre for Information Resilience (CIR) launched the Eyes on Russia (EoR)\footnote{\url{https://eyesonrussia.org/}} project in January 2022 with the aim of gathering and verifying media related to Russia's invasion of Ukraine. The project's primary objective is to provide access to verified information through a database and an interactive map, benefiting journalists, non-governmental organizations (NGOs), policymakers, and the public. The interactive map displays relevant information such as the data source, a description of the event, location coordinates, and the extent of damage caused. Furthermore, it includes a variety of properties, including the country name, province, city, coordinates, date, damage level, and source of information. 
% Civilian Harm:
The Civilian Harm in Ukraine TimeMap (CH)\footnote{\url{https://ukraine.bellingcat.com/}} is a similar project that provides a comprehensive record of such incidents by including source links, precise location data determined by the Global Authentication Project and Bellingcat researchers, and a brief description based on visual evidence. Its structured data can be used for further analysis and research to understand better the impact of the conflict on civilians in Ukraine.
% a summary
EoR and CH are the two datasets selected for this study given their similar approach in generating and representing events. Both projects focus on damage reporting and serve as important resources for those seeking accurate and verified information to aid decisions for the resilience of Ukraine.

% motivation
Accurate and complete documentation of the damage could benefit projects for resilience in multiple ways. Linked Data is structured data that can be interlinked with other data, which enables additional functions through semantic queries. While it is not common to use Linked Data in projects for resilience, past projects have demonstrated the use of Linked Data in decision-making in government, NGOs, and societal organizations. For example, the Brazilian government used ontologies and enriched data from various governments, resulting in a DBpedia-like Government Open Linked Data - DBGOldBr \cite{victorino2018transforming}. Our project explores how the transformation of event data into Linked Data facilitates an integrated and more complete description of events. For example, we consider the following event\footnote{The event was extracted manually based on the Twitter post \url{https://twitter.com/KyivIndependent/status/1501218105342763020}.} in CH that happened on 7th March 2022. It was reported to have ``Hospital destroyed by explosion''. Its location information is ``Izum, Kharkiv region'' and it lacks information about the postal code.

% ontologies can serve as a unified vocabulary for the representation of such knowledge.
% ad-hoc 
While the above-mentioned datasets were initially designed for their platforms, one can take advantage of ontologies and Linked Data technologies to provide a unique representation of entities such as cities and provinces to reduce ambiguity, which could make it easier for integration and verification, and enable interoperability with datasets in other disciplines (e.g. economic and social-/historical- data). In this paper, we attempt to convert and unify structured geo-annotated datasets. More specifically, we convert two existing geo-annotated datasets dedicated to damage reporting in Ukraine to their corresponding representation as Linked Data. We propose a pipeline for the integration of datasets and demonstrate the use of the resulting dataset. Finally, we evaluate the quality of the integrated data and demonstrate its application by several use cases. 

Our research question is: \textbf{How to unify geo-annotated events in multiple datasets about damaging events?} We answer this question by studying the following sub-research questions:

\begin{description}
  % \item[SRQ1:] How can we provide a unified representation and deal with regional identification?
  \item[SRQ1:] How can we provide a unified representation of information in the datasets as Linked Data?
  \item[SRQ2:] How can we enrich the converted Linked Data with geospatial information?
  \item[SRQ3:] How can we integrate datasets by identifying and merging entities that describe the same events?
  \item[SRQ4:] What is the quality of the resulting unified data?

\end{description}

% Research output
% The research output of this paper includes 1) the converted datasets together with related resources; 2) an integrated dataset; 3) a pipeline with open source code that can be adapted to future datasets; 4) use cases with SPARQL queries.

% Research output
The research output of the paper includes the converted dataset together with related resources, an integrated dataset, a reusable pipeline with open source code that can be adapted to future datasets, the mapping between entities in the datasets, and some use cases with their SPARQL queries.

This paper is organized as follows. Section \ref{section:related-work} investigates related approaches to the representation of damaging and resilience information. Section \ref{section:data-processing} compares datasets from various resources and selects the datasets to be studied in this paper based on criteria. Section \ref{sub-section:conversion} explains the conversion process then in  Sections \ref{sub-section:enrichment}  we include details of data enrichment and multilingual representation. Following that, we convert the selected datasets to linked data and integrate them in Section \ref{section:integration}. Furthermore, we indicate the evaluation process in Section \ref{section:evaluation-publication}. To demonstrate the use of our integrated dataset, we designed some use cases in Section \ref{section:use-cases}. Finally, we discuss the limitations of our approach, alternative approaches, as well as future work in Section \ref{section:discussions-conclusion}.

\section{Related Work}
\label{section:related-work}
Various open datasets provide valuable insights into the Russo-Ukrainian conflict from different perspectives. Two prominent datasets in this domain are WikiEvents \cite{wikievents} and ACLED \cite{acled}. WikiEvents is a collection of entries automatically curated based on Wikipedia's Current Events portal.\footnote{\url{https://en.wikipedia.org/wiki/Portal:Current_events}} Its NLP downstream pipeline extracts 21,275 events including around a thousand events related to the Russian invasion of Ukraine. The dataset enables researchers and analysts to gain insights into the complex situation, contributing to the development of effective strategies to protect civilians, promote peace, and estimate resources for resilience projects. ACLED, on the other hand, is a much larger dataset\footnote{\url{https://acleddata.com/2023/03/01/war-in-ukraine-one-year-on-nowhere-safe/}, visited on 13th June, 2023.} with over one million events, including approximately 40,000 political violence events across Ukraine. Although ACLED provides a vast amount of data, it primarily focuses on military aspects. A significant portion of its events is centered around shelling, artillery, and missile strikes. In response to the conflict, the Centre for Information Resilience (CIR) \cite{cir} launched the Eyes on Russia (EoR)\footnote{\url{https://eyesonrussia.org/}} project in January 2022. EoR aims to gather and verify media content related to Russia's invasion of Ukraine, providing access to verified information through a database and an interactive map. This project is valuable for journalists, NGOs, policymakers, and the public, offering insights into the conflict and its impact. Another noteworthy project is the Civilian Harm in Ukraine TimeMap(CH)\footnote{\url{https://ukraine.bellingcat.com/}}, which provides a descriptive record of incidents, including source links, precise location data, and descriptions based on visual evidence. The structured data provided by CH facilitates a better understanding of the impact of the conflict on Ukrainian civilians.
% While these datasets were initially designed for their respective platforms, our project targets to leverage ontologies and linked data technologies to provide a unified representation of entities such as cities and provinces, reducing ambiguity and enabling easier integration and interoperability with datasets in other disciplines \cite{SmartCityData}. By reviewing the existing literature and datasets related to damage reporting and resilience in Ukraine, we can identify gaps and opportunities for our research, which focuses on the conversion, unification, and enrichment of geo-annotated datasets to facilitate integrated and more complete event descriptions. In addition to the aforementioned datasets, several studies have explored the use of linked open data and ethical considerations in crisis events and big data environments.
One relevant article focuses on transforming open data into linked open data using ontologies in the context of the Brazilian government's DBpedia-like Government Open Linked Data (DBGOldBr)\cite{victorino2018transforming}. initiative. The authors highlight the challenges faced by the Brazilian government in organizing and providing access to vast amounts of data. They explain how the use of linked open data and ontologies can help overcome these challenges by improving data integration, interoperability, and reuse. The article describes the ontology development process and provides examples of how DBGOldBr was utilized to link and enrich data from different government agencies.

Another study addresses the limitations of using social media and phone data to understand people's responses during crises \cite{limits-crisis-data}. Using examples of Hurricane Sandy and the Haiti Earthquake, the authors demonstrate how relying solely on platforms like Twitter and crowdsourced text messages raises questions about the representativeness of the data. The article emphasizes the importance of being aware of ethical frameworks and understanding the limitations of the data to obtain a more accurate understanding of crisis events. Furthermore, it highlights the need for researchers to consider the impact of their work on privacy and consent.

Ethical considerations in using machine learning models on social media data are explored in a separate paper \cite{kraft2022ethical}. The report provides an overview of social media's evolution and discusses the various ethical risks associated with employing machine learning techniques for data collection. Privacy concerns, bias in algorithms, and the manipulation of information are among the ethical challenges discussed. The study emphasizes the responsible collection, use, and sharing of data, as well as the importance of transparent and fair decision-making processes.

These studies could be enhanced by incorporating linked data approaches. By leveraging linked data and addressing ethical concerns, researchers can improve the integration, analysis, and interpretation of data in the context of resilience projects, leading to more informed decision-making processes.

\section{Data Selection} %Shuai: one-two pages and leave the rest to the appendix
\label{section:data-processing}
% Survey+ some analysis.
% EyesOnRussia:
% The Centre for Information Resilience (CIR) launched the Eyes on Russia project in January 2022 with the aim of gathering and verifying media related to Russia's invasion of Ukraine. The project's primary objective is to provide access to verified information through a database and an interactive map, benefiting journalists, non-governmental organizations (NGOs), policymakers, and the public. The interactive map displays relevant information such as the data source, a description of the event, location coordinates, and the extent of damage caused.
% Furthermore, the 

Next, we present details of representative datasets about damaging events in Ukraine and consider their feasibility for our linked-data approach. 

EyesOnRussia:
The EyesOnRussia project has generated a dataset that includes a variety of properties, including the country name, province, city, coordinates, date, damage level, and source of information. This dataset is provided in JSON format, and it is intended to support further analysis and research on the conflict between Russia and Ukraine.
Overall, the Eyes on Russia project serves as an important resource for those seeking accurate and verified information about the ongoing conflict, and its availability in a structured dataset further enables researchers to gain insights into the conflict's complexities.
 \footnote{Eyes on Russia dataset was retrieved on 30th of April 2023}
 
% Civilian Harm:
% The Civilian Harm in Ukraine project is focused on gathering and documenting incidents that may have resulted in harm to civilians in the country. These incidents include attacks on civilian areas and damage to civilian infrastructure. The project aims to provide a comprehensive record of such incidents by including source links, precise location data determined by the Global Authentication Project and Bellingcat researchers, and a brief description based on visual evidence.
Civilian Harm: 
The Civilian Harm dataset contains several properties, including city, coordinates, date, and source of information, which are available in JSON or CSV format. This structured data can be used for further analysis and research to understand better the impact of the conflict on civilians in Ukraine.
Overall, the Civilian Harm in Ukraine project plays an essential role in documenting and raising awareness of the potential harm to civilians in the ongoing conflict in the country. The project's structured data further enables researchers and analysts to gain insights into the complexities of the situation, ultimately contributing to the development of effective strategies to protect civilians and promote peace.
\footnote{Civilian Harm dataset was retrieved on 30th of April 2023}

leedrake5-Russia-Ukraine\cite{leedrake5}:
The Russia-Ukraine War tracker is a tool that leverages data from Oryx's site to visualize equipment losses since Russia's invasion of Ukraine on February 24th. This tracker provides independent verification of the destroyed vehicles and equipment for which photo or videographic evidence is available. The data is sourced from a public Google sheet that is updated based on the latest information available for each day.
The properties of the Russia-Ukraine War tracker include unit loss by country and date, which are presented in the Google Sheets format. This structured data allows for the analysis of the losses incurred by various military units over time and provides insights into the ongoing conflict between Russia and Ukraine.
Overall, the Russia-Ukraine War tracker serves as a useful resource for those seeking to understand the impact of the conflict on military equipment and operations. The availability of structured data in the Google Sheets format further facilitates the analysis of losses incurred by various military units and can help inform strategic decision-making.

StandWithUkraine\cite{standwithukraine_2023}:
StandWithUkraine is an open data service that provides free and easy access to information about Ukraine. The service offers a simple interface and an API for developers to access the data, making it a valuable resource for researchers, analysts, and other potential users interested in understanding various aspects of Ukraine.
The Ukrainian Data Hub is part of StandWithUkraine, and provides a repository of structured data on various topics, including demographics, economics, and social issues. This data can be used to identify trends, develop insights, and support evidence-based decision-making. Additionally, StandWithUkraine maintains a GitHub repository with statistics and information about humanitarian numbers, providing a valuable resource for those interested in the humanitarian situation in Ukraine. Overall, StandWithUkraine plays an important role in promoting open data and transparency in Ukraine, making valuable information accessible to a wide range of users. The service's simple interface and free access via API and GitHub repository further facilitate the use of data to support research and analysis.

Russia block list\cite{veelenga_2023}: 
The Russia block list is a compilation of companies that have suspended their operations in Russia due to various reasons. These reasons may include economic sanctions, political instability, or other factors that have made it difficult for these companies to continue operating in Russia.
The block list provides a valuable resource for those interested in understanding the impact of economic and political factors on businesses operating in Russia. By compiling information about companies that have suspended their operations, the block list can help identify patterns and trends in the business environment in Russia and inform strategic decision-making.
Overall, the Russia block list serves as an important tool for researchers, analysts, and other stakeholders interested in understanding the impact of various factors on businesses operating in Russia. By providing a list of companies that have suspended their operations, the block list can help shed light on the challenges and opportunities facing businesses in this complex and dynamic environment.

ACLED: 
The ACLED project collects data on various forms of political violence, including armed conflict, riots, protests, and non-violent campaigns, among others. The data includes information on the location, date, actors involved, and fatalities and injuries resulting from the event. The project also collects data on the type of violence used, such as bombings, assassinations, and torture, as well as the targets of the violence, such as civilians, military personnel, or political leaders. Moreover, it provides properties such as country name, province, city, coordinates, date, and the source of information.

% See Appendix X for details of the survey and scope of voulunteers. 

The criteria for inclusion and exclusion are defined with a main focus on describing damaging events happening in Ukraine and the richness and reliability of information to be converted to linked data. Therefore, the datasets we choose to include are preferred to have at least some of the following information (but not necessarily all entries): some description of the event,
the geospatial location of the event (i.e. the coordinates), the source of the event,
the date of the event, and the availability of datasets (so it is possible to convert the open data to linked open data). Based on the defined criteria, the EyesOnRussia and Civilian Harm datasets meet the requirement. Additionally, we will retrieve address information such as province and postal code using coordinates via geocoding services. It is also possible to obtain the geocoding by converting a location's coordinates (latitude and longitude) into a human-readable address. Moreover, the two datasets selected have some overlap in information, which could be used for semi-automatic validity checking at a later stage. Using the criteria above, Leedrake5-Russia-Ukraine and StandWithUkraine datasets will be excluded, although they provide such interesting facts and information on the war in Ukraine, these datasets are beyond the scope of the project as they do not provide the required description of the events, nor a detailed time and location. Instead, these datasets provide statistical data including humanitarian and refugee records that may be integrated in future work. Similarly, ACLED contains mostly data about military activities, violence, and explosions, which are related but not exactly the kind of reports of damage we prefer and will be left for future work. 

\section{Data Conversion}
\label{sub-section:conversion}
Our examination of the datasets shows that the fields and formats of reported events can vary significantly. This is partially due to the lack of use of controlled vocabulary and ontology. Take the location information of CH for example, the event in Section \ref{section:introduction} has location information ``Izum, Kharkiv region''. However, we observed other formats such as ``Kharkiv'', ``Merefa, Kharkiv'', as well as poorly formatted strings such as ``$\backslash$r$\backslash$nZhytomyr'', and mistakes such as ``Kyiv region, Donetsk''. To answer SRQ1, we select entities and relations from popular ontologies such as \texttt{schema.org}\footnote{\url{https://schema.org/docs/schemas.html}}, the Dublin Core\footnote{\url{https://www.dublincore.org/}}, Simple Event Ontology\footnote{\url{https://semanticweb.cs.vu.nl/2009/11/sem/}}, and the GeoNames\footnote{\url{https://www.geonames.org/}} for a unique representation of (geo-)information of events.
% To answer SRQ1, we select entities and relations from popular ontologies such as \texttt{schema.org} \cite{guha2016schema}, the Dublin Core \cite{weibel2000dublin}, Simple Event Ontology \cite{simpleEventOntology}, and the GeoNames \cite{maltese2013semantic} for a unique representation of (geo-)information of events. 
In addition, we also introduce some relations in our own namespace. Moreover, some specific information is not generic between datasets, e.g. violence level and type of damage `Civilian Infrastructure Damage'. We include such information in the comment (as the object of \texttt{rdfs:comment}) to be studied in future work. 

We assign a Uniform Resource Identifier (URI) to each event. We model that each event is of type \textit{Event} as in the Simple Event Ontology \cite{simpleEventOntology}. We noticed that many events were reported with an accurate date but not the exact time. In fact, many happened at exactly 00:00:00, which could be the default time setting. Therefore, we ignore the exact time of the event and take the day without the time. Following that, we use its coordinates and find its unique representation of province, city, and postal code in GeoNames. As for the example in section \ref{section:introduction}, the reported province/region is Kharkiv. We retrieve Kharkiv's corresponding URI in GeoNames: \url{http://sws.geonames.org/706483/}. However, its postal code is still missing. This leads to the step of data enrichment in the next section.

% 

% \begin{table}[!ht]
% \begin{tabular}{l|ll|ll|ll|ll|ll}
% \multirow{2}{*}{} & \multicolumn{2}{c|}{EoR} & \multicolumn{2}{c|}{CH} & \multicolumn{2}{c|}{EoR'} & \multicolumn{2}{c|}{CH'} & \multicolumn{2}{c}{Integrated} \\ 
%  & |I| & |E| & |I| & |E|  & |I| & |E|   & |I| & |E|  & |I| & |E|  \\ \hline
% city &  &  &  &  &  &  &  &  &  &  \\
% city &  &  &  &  &  &  &  &  &  &  \\
% city &  &  &  &  &  &  &  &  &  &  \\
% city &  &  &  &  &  &  &  &  &  &  \\
% city &  &  &  &  &  &  &  &  &  &  \\
% \begin{tabular}[c]{@{}l@{}}damage\\ level\end{tabular}&  &  &  &  &  &  &  &  &  &  \\ \hline
% \begin{tabular}[c]{@{}l@{}}overall\\ entries\end{tabular} &  &  &  &  &  &  &  &  &  & 
% \end{tabular}
% \label{tab:overview}
% \end{table}

\begin{table*}[!ht]
% \footnotesize
\centering
\begin{tabular}{l|llp{3.7cm}|llp{3.7cm}}
 & \multicolumn{3}{c|}{\textbf{EoR}} & \multicolumn{3}{c}{\textbf{CH}} \\
 & \textbf{O} & \textbf{CE}    & \textbf{comment} & \textbf{O} &  \textbf{CE} &\textbf{comment} \\ \hline
country & 9308 & 9308  & obtained GeoNames' country URI using the string  & 0  & 1105 & obtained GeoNames' country URI using the coordinates \\
city & 9308 & 9308  & obtained GeoNames' city URI using the string and coordinates & unknown & 1105 &converted from string to GeoNames' city URI or retrieved using coordinates  \\
province & 9308 & 9308  & 25 were manually corrected due to incorrect spelling  & unknown & 1105   & for inconsistent representation, their province was obtained as GeoNames' province URI by using their coordinates  \\
date & 9308 & 9308  & converted from string to date:xsd format & 1105  & 1105 & converted from string to date:xsd  \\
coordinates & 9308 & 9308  & added as GeoCoordinates format  & 1105  & 1105 & added as GeoCoordinates format  \\
postal code & 0 & 9223  & retrieved from GeoNames using the coordinates (85 entries do not have a corresponding postal code in GeoNames) & 0  & 1105 & retrieved from GeoNames using the coordinates \\
description & 9306 & 9306  & two events lack description.  & 1105  & 1105 & kept original  \\
URL & 9308 & 9308  &   & 1057  & 1057 &  \\
violence level & 9296 & 0  &  the violence level was left as comments due to lack of standards and definition & 0  & 0 & CH does not have the violence level  \\ \hline 
% postcode & 9252 & 9308  & Only 8884 cities found in GeoNames. 368 were about villages, towns, and other names that do not exist in GeoNames' cities. 56 events have no corresponding information. We retrieved this information from their coordinates in GeoNames. & 9991 & 9991   & Converted from string to country URI from GeoNames  \\
% province & X & O & Converted from string to Province URI from GeoNames  & - & + & Converted from string to Province URI from GeoNames  \\
% city & X & O & Converted from string to City URI from GeoNames  & X & O & Converted from string to City URI from GeoNames \\
% postcode & - & + & Retrieved via Geonames web services & - & + & Retrieved via Geonames web services \\
% street & X & - & Removed because of low availability & - & - & Unchanged \\
% location & X & X & Unchanged & X & X & Unchanged \\
% date/time & X & O & Removed because of low availability & - & - & Unchanged \\
% violence level & X & - &  & ? & O & 
\#events & 9308 & 9308  &   & 1105  & 1105 & 
\end{tabular}
\label{table:convert}
\caption{Comparison of old vs new Eyes on Russia and Civilian harm datasets entries (O: The original dataset, CE: the dataset after conversion and enrichment.)}
% \medskip
% \small

% X standards for entries that were in the old dataset in the original format; O standards for entries that were there before but their format changed in the new dataset; + indicates that additional information can be obtained and was added to the dataset; - indicates that this entries are missing or got removed
\end{table*}

\section{Data Enrichment}
\label{sub-section:enrichment}

It was noticed that some information is not explicitly provided but can be inferred. For example, the postal code can be retrieved by calling GeoNames' APIs. Recall our example in Section \ref{section:introduction}, the missing information postal code is 64305. Not all information was represented correctly. Take EoR for example,  only 8884 events have their city information formatted correctly and found in GeoNames. Another 368 associated strings were about villages, towns, local neighborhoods, or other names that do not exist as cities using GeoNames. 56 events have none of the corresponding information mentioned above. Therefore, we retrieved this information from their coordinates in GeoNames. Difficulty due to spelling errors and multilingual cases were manually resolved. Table 1 presents a summary of conversion and enrichment.

\begin{figure}[!ht]
  \centering
  \includegraphics[width=9cm]{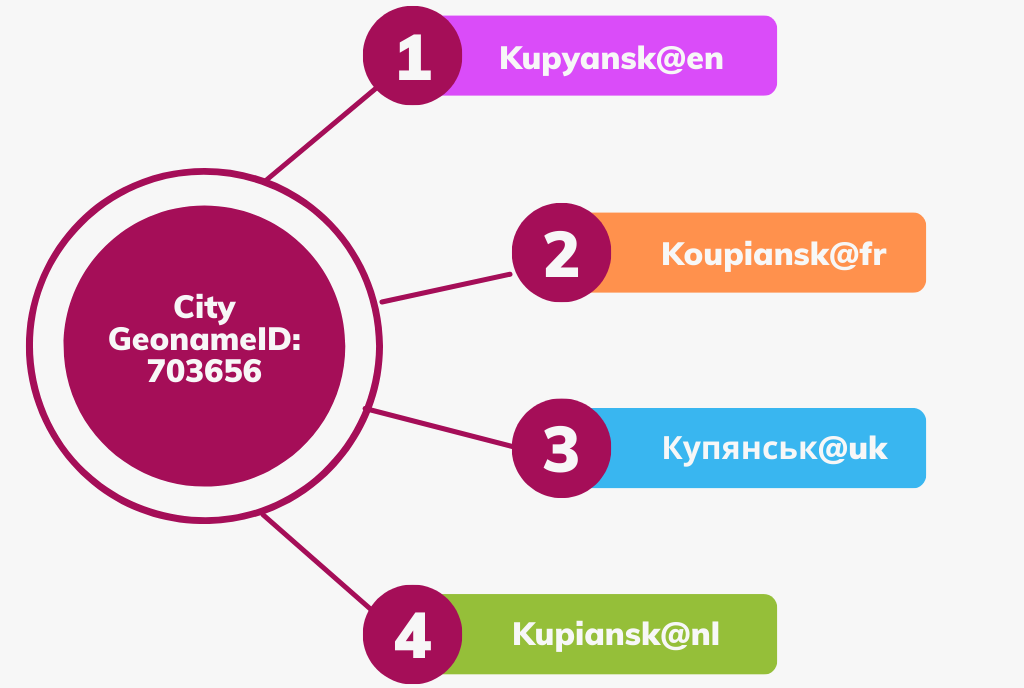}
  \caption{Multilingual representation of the Kupyansk City}
  \label{fig:multiligual}
\end{figure}

Multilingual information is vital to accommodate collaboration across countries, improve usability for users, and enhance interoperability. The inclusion of multiple languages empowers diverse users to engage with and enhances accessibility, compatibility, and supports diverse resilience strategies that overcome linguistic obstacles. This inclusive approach promotes resilience and facilitates knowledge sharing among global communities \cite{smith2022role}. Figure \ref{fig:multiligual} displays multilingual labels for the city of Kupyansk, showcasing the diverse linguistic representation in our data.

% 2. How you got the data. 

To achieve the enrichment of data with multilingual labels, the Geonames web service API was employed. The process involved the following steps. Firstly, the geonameID obtained previously was utilized as a reference. Secondly, requests were made to the Geonames API, specifying the desired languages (such as French, English, Ukrainian, and Dutch) in the API calls. The API responded with the corresponding city names in the requested languages. These multilingual city names were then saved in a JSON file for use during data conversion to RDF triples. Finally, the saved labels were incorporated into the dataset, associating each city with its respective multilingual names. This enabled the data to be enriched with multilingual labels, providing language-specific information for further analysis.

\section{Data Integration}
\label{section:integration}

% That algorithm in text:
% A summary of the algorithm. 
Next, we study SRQ3. Our manual examination shows that cases where one event was reported two or multiple times are very rare. Therefore, we rely on the Unique Name Assumption for both datasets: no event was reported twice at a close distance in the same dataset. Algorithm \ref{algo:integration} takes into consideration the distance of events from two datasets and their description. We manually fine-tuned all the parameters. The output of the following algorithm consists of 1) pairs of events that we consider potentially identical (denoted $S$) and 2) pairs of events that are close to each other but not identical (denoted $T$).\footnote{Other pairs of events are stored for manual examination in future work.} As for string similarity, we took advantage of the \textit{SequenceMatcher} function in the \textit{difflib} Python package.\footnote{https://docs.python.org/3/library/difflib.html} Other sequence comparison methods will be explored in the future. 

\begin{algorithm}[!ht]

\For{ each pair of events $(i, j)$ with identical city and date}{
$d \gets$ the distance between $i$ and $j$;

$s \gets$ the similarity between the description of $i$ and $j$;

\eIf{$i$ and $j$ are backed by the same social media link and $s > 0.55$ and $d < 2 (km)$}
{
    add $(i, j)$ to $S$
}{
    \If{`area' is in the description of $i$ or $j$}
    {
        \eIf{$s > 0.75$ and $d < 2 (km)$}{
        add $(i, j)$ to $S$
        }{
        add $(i, j)$ to $T$
        }
    }

   \If{keywords such as `school', `hospital' are in the description of $i$ or $j$}
    {
        \eIf{$s > 0.55$ and $d < 1 (km)$}{
        add $(i, j)$ to $S$
        }{
        add $(i, j)$ to $T$
        }
    }
    }
    
}

\caption{Data integration using distance, description, and associated  link to social media content}
\label{algo:integration}
 \end{algorithm}

% \begin{figure}[h]
%   \centering
%   \includegraphics[width=0.65\linewidth]{hasSource.png}
%   \caption{Integration of events}
%   \label{fig:integration}
% \end{figure}
% explain the algorithm above
Our manual examination shows that the coordinates of reported identical events about an `area' could be some distance apart. Therefore, we take two different strategies for areas and other cases separately. We consider a broader radius of 2km for events about `area'. For other cases, we consider only the keywords about theater, church, school, hospital, building, house, flat, station, etc. Since our dataset focuses mostly on damage reporting, other reported events such as military operations are not considered for merging and will be filtered out in the future.
% Merging the two datasets:
We identified 206 pairs of events. For each pair, we associate them with a new event URI that represents the integration of these two events. Moreover, we introduce an additional \texttt{hasPrimarySource} relation in our namespace for the primary source of the event (the event with richer information). Recall that there are 9,308 and 1,105 events for EoR and CH, respectively. Overall, the integrated dataset includes 10,207 events.

% (see Figure \ref{fig:integration}). 

% Merging the two triples files into one 
% Dealing with inconsistency: if two events have the same data, postalCode and city name, they are considered Duplicates, and they should be removed
% Dealing with similarities: if two or more events have the same city, postal code and date but a similar description they should be merged and the description is extended 
% Leaving dissimilar events: if two or more events have the same city, postal code and date but a totally different description they should be kept and treated as separate.

% Provenence information. 

% After merging identical triples, the integrated dataset consists of Y entries in X triples (see Table \ref{tab:overview}).

% links to DBpedia
% In addition, we publish X identity links () to DBpedia. This gives us the possibility to retrieve information for better illustration and use cases.  

\section{Data Evaluation and Data Publication}
\label{section:evaluation-publication}

Finally, for SR4, we assess the quality of our algorithm and the resulting datasets. For the former, we created a questionnaire that consists of randomly selected 10 pairs of events from $S$ (pairs of events considered identical) with 10 additional pairs of events selected from $T$ (pairs of events from the same city, on the same day, and close to each other but not considered identical). We received 6 valid submissions by the deadline.\footnote{The questionnaire and results are included in the supplementary material.} By assigning a number to each answer\footnote{2 for `Very likely', 1 for `Likely', 0 for `Unsure', -1 for `Unlikely', and -2 for `Very unlikely'.}, our analysis of the results indicates that pairs of events considered identical by our algorithm have an average of 1.38 (between `Likely' and `Very likely' to be identical). In comparison, that of other events is -0.45 (between `Unsure' and `Unlikely'). This shows that our algorithm has good precision while those we decide to leave out remain unsure. 

To prevent unintended use, the integrated dataset is only available upon request. The rest of our datasets are hosted on the TriplyDB platform.\footnote{\url{https://triplydb.com/linked4resilience/}. The converted datasets of CH and EoR can be found but the integrated dataset is accessible upon request.} Passing it through TriplyDB's data processing pipeline ensures syntactic correctness. As sanity checks of our data, we run SPARQL queries to manually validate the ranges of data points on temporal and spatial dimensions. The SPARQL queries, the converted datasets, the code, the questionnaire, as well as other supplementary materials, are available on GitHub.\footnote{\url{https://github.com/LinkedData4Resilience/linked-data}.} A demo and other related resources can be found on the Linked4Resilience project website: \url{https://linked4resilience.eu/}.

\section{Use Cases}
\label{section:use-cases}

In this section, we demonstrate the practical application of our data by designing SPARQL queries corresponding to some resilience aspects. Use case \ref{section:usecase1} highlights the value of visualizing events in Kherson on a map, and Use case \ref{section:usecase2} analyzes the multilingual representation of labels in the integrated dataset. Additionally, use case \ref{section:usecase3} showcases how to gain insights from timelapse visualizations of events about public facilities, and use case \ref{section:usecase4} gives the analysis of monthly attacked regions. Furthermore, use case \ref{section:usecase5} is a proof-of-concept study that indicates the examination of the ratio of children's deaths to monthly attacks offers valuable insights into the humanitarian impact of the conflict. Finally, use case \ref{section:usecase6} attempts to find locations without shelters nearby. Together, these use cases illustrate the diverse applications of the integration of data to describe the various dimensions of the conflict in Ukraine.

% Section \ref{subsect:usecase1}.1 gives a  use case that generates a map displaying the events that occurred in Kherson as small blue circles. The remaining four use cases deal with statistical aspects of resilience

%Shuai: You don't need to explain anything here. Just include the SPARQL queries in each use case
% \section{SPARQL Query}
% \label{section:sparql}
% % example of Geosparql

% \section{Use Cases}
% \label{section:use-cases}

\subsection{Use Case 1: Damage visualization}
\label{section:usecase1}
\begin{figure}[ht]
  \centering
  \includegraphics[width=0.78\linewidth]{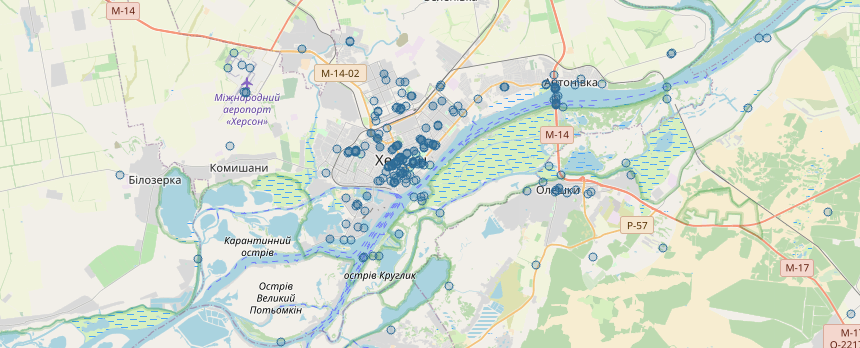}
  \caption{A visual representation of the damaging events in Kherson}
  \label{fig:Events in Kherson}
\end{figure}

As a demonstration\footnote{A demo showing the use of the datasets at \url{https://youtu.be/E_fr1KzfsVs?feature=shared}.} of the use of our integrated dataset, Figure \ref{fig:Events in Kherson} showcases the outcome of a SPARQL query that retrieves events in Kherson within the integrated datasets from 1st October 2022 to 28th February 2023, which are then visualized on a map. By mapping these events, it becomes possible to identify patterns, concentrations, and trends on the map. The visual representation aids researchers and policymakers in gaining valuable insights into the spatial distribution of these events, thus facilitating better understanding and analysis.

This use case exemplifies the practical application of integrated datasets and the geographic visualization.
The SPARQL query \ref{lst:sparql1} aims to retrieve the location of the events and display them on the map using the primary sources predict, location, date, and geospatial coordinates within a specified timeframe. The query generates a WKT representation to support the visualization, and integration of the retrieved data.

\begin{lstlisting}[captionpos=b, caption=SPARQL query for damaging events map, label=lst:sparql1,
   basicstyle=\small,frame=single]
PREFIX sdo: <https://schema.org/>
PREFIX rdfs: <http://www.w3.org/2000/01/rdf-schema#>
PREFIX rdf: <http://www.w3.org/1999/02/22-rdf-syntax-ns#>
PREFIX xsd: <http://www.w3.org/2001/XMLSchema#>
PREFIX purl: <http://purl.org/dc/terms/>

SELECT ?wkt WHERE {
  ?aggregateEvent <https://linked4resilience.eu/ontology/hasPrimarySource> ?event .
  ?event sdo:location ?loc.
  ?event purl:date ?date .
  ?loc sdo:geo ?geo.
  ?geo sdo:latitude ?lat .
  ?geo sdo:longitude ?lng .
  filter(?startDate <= ?date)
  filter(?date <= ?endDate)
  bind(
    strdt(
      concat("POINT(",str(?lng)," ",str(?lat),")") ,
      <http://www.opengis.net/ont/geosparql#wktLiteral>
    )
      as ?wkt)
} 
\end{lstlisting}

\subsection{Use Case 2: Public facilities damaged}
\label{section:usecase2}

\begin{figure}[ht]
 \centering\includegraphics[width=0.65\linewidth]{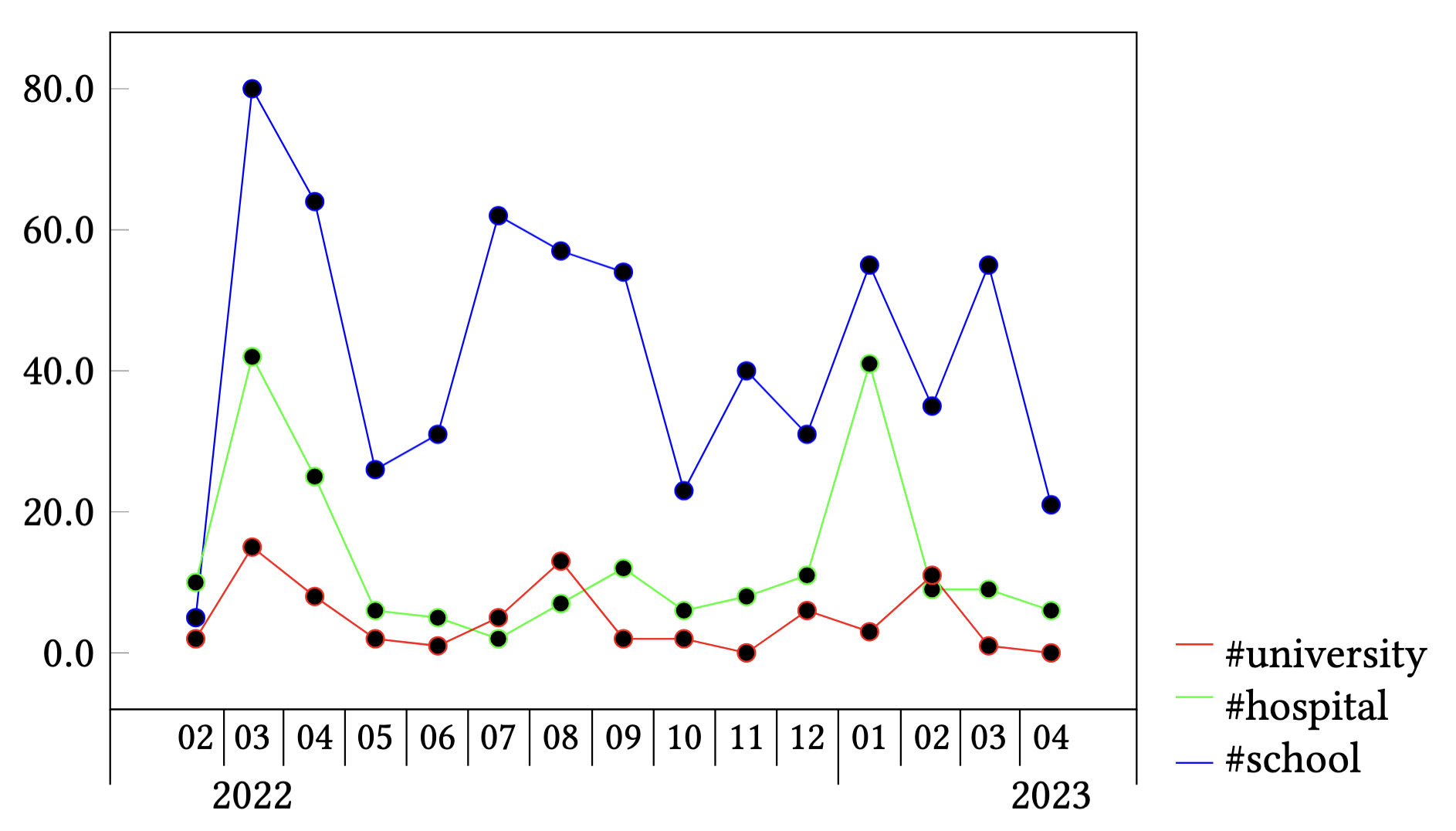}
    \caption{Timelapse of events about public facilities}
    \label{fig:time_plot}
\end{figure}

We dedicate this use case to showcase the timelapse of damaging events specifically related to schools and hospitals. Figure \ref{fig:time_plot} presents a visual representation of the dates and the corresponding number of events concerning schools, universities, and hospitals during the period from 1st February 2022 to 30th April 2023. This timelapse provides an overview of the frequency and distribution of damaging events targeting educational and healthcare facilities over the specified timeframe. By observing the patterns and fluctuations depicted in the visualization, we can gain insights into the impact and severity of the conflict on these vital institutions. SPARQL query \ref{lst:sparql2} aims to retrieve and calculate the maximum number of events that occurred in each month within a specified date range using a filter for a keyword. The query provides focused insights into event frequencies and patterns.
% The following code is adapted from  https://tex.stackexchange.com/questions/268994/pgfplots-graph-labels-year-month

\begin{lstlisting}[captionpos=b, caption=SPARQL query for keyword damaged events time series , label=lst:sparql2,
   basicstyle=\small ,frame=single]
prefix dct: <http://purl.org/dc/terms/>

select ?monthyear (max(?numEvents) as ?numevts) where {
  {
    values ?monthyear {
      "2022-02" "2022-03" "2022-04" "2022-05" "2022-06" "2022-07"
      "2022-08" "2022-09" "2022-10" "2022-11" "2022-12" "2023-01"
      "2023-02" "2023-03" "2023-04"
    }
    bind(0 as ?numEvents)
  }union
  {
    select (count(distinct ?aggregateEvent) as ?numEvents) ?monthyear where {
      values ?monthyear {
        "2022-02" "2022-03" "2022-04" "2022-05" "2022-06" "2022-07"
        "2022-08" "2022-09" "2022-10" "2022-11" "2022-12" "2023-01"
        "2023-02" "2023-03" "2023-04"
      }
      ?aggregateEvent <https://linked4resilience.eu/ontology/hasPrimarySource> ?event .
      ?event ?predicate ?object .
      filter(isliteral(?object)) .
      filter(contains(lcase(str(?object)),lcase(?searchTerm)))
      ?event dct:date ?time .
      bind(replace(replace(replace(concat('0',str(month(?time))),
      '010','10'),'011','11'),'012','12') as ?monthstr)
      bind (concat(year(?time), '-', ?monthstr) as ?computed_monthyear)
      filter(?computed_monthyear = ?monthyear)
    }
  }
} order by ?monthyear
\end{lstlisting}

% \subsection*{Use Case 3: Damage estimation}
% \label{subsect:usecase3}

% \label{subsect:usecase2}

% 

% In Kharkiv, the war has resulted in the deaths of hundreds of civilians, large-scale destruction of infrastructure, and also a humanitarian crisis, which is getting worse every day. There are estimates that more than 1,000 buildings have been destroyed, of which more than 700 are multi-storey apartment buildings, which are no longer habitable.
% \cite{Chumachenkoo796}

\subsection{Use Case 3: Multilingual representation of labels}
\label{section:usecase3}
\begin{figure}[ht]
  \centering
  \includegraphics[width=0.92\linewidth]{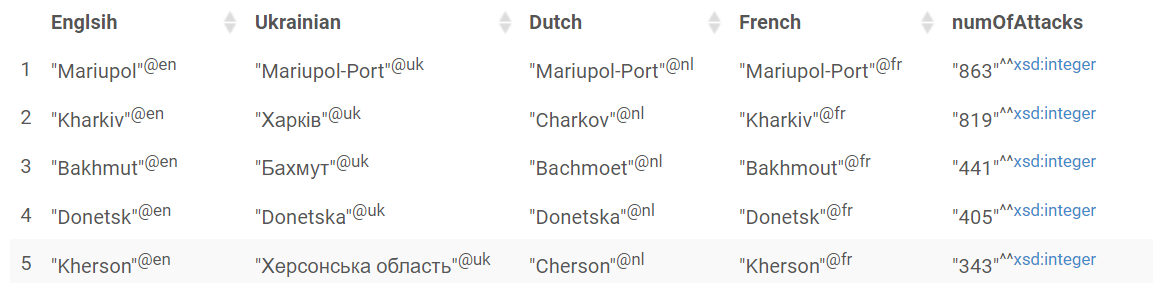}
  \caption{YASGUI displaying the top 5 attacked cities in various languages between February 2022 and April 2023}
  \label{fig:attacked-cities-mulitlingual}
\end{figure}

Incorporating multilingual information in a resilience project utilizing linked data is crucial for effective international collaboration, enhancing usability for many users, and improving interoperability. This holistic approach ensures that potential users of the data can actively participate, comprehend, and contribute to the project, ultimately fostering more resilient and inclusive communities.
Figure \ref{fig:attacked-cities-mulitlingual} displays multilingual labels for the city of Kupyansk, showcasing the diverse linguistic representation in our data. SPARQL query \ref{lst:sparql3} aims to retrieve distinct city name variations in English, Ukrainian, Dutch, and French. The query applies language filters, counts the occurrences of city name combinations, groups the results, and presents the top five combinations with the highest occurrence count.

% sec:multilingual

\begin{lstlisting}[captionpos=b, caption=SPARQL query for representing cities in multilingual labels, label=lst:sparql3,
   basicstyle=\small ,frame=single]
PREFIX xsd: <http://www.w3.org/2001/XMLSchema#>
PREFIX rdf: <http://www.w3.org/1999/02/22-rdf-syntax-ns#>
PREFIX rdfs: <http://www.w3.org/2000/01/rdf-schema#>
PREFIX l4r_o: <https://linked4resilience.eu/ontology/>

SELECT distinct ?English ?Ukrainian ?Dutch ?French(count(?sub) as ?numOccurences) WHERE {
  ?sub l4r_o:cityName ?English .
  ?sub l4r_o:cityName ?Ukrainian .
  ?sub l4r_o:cityName ?Dutch .
  ?sub l4r_o:cityName ?French .
  filter(LANG(?English) = "en")
  filter(LANG(?Ukrainian) = "uk")
  filter(LANG(?Dutch) = "nl")
  filter(LANG(?French) = "fr")
} group by ?english ?Ukrainian ?Dutch ?French order by desc(?numOccurences)
LIMIT 5
\end{lstlisting}

\subsection{Use Case 4: Monthly most attacked regions}
\label{section:usecase4}
\begin{figure}[ht]
  \centering
  \includegraphics[width=0.95\linewidth]{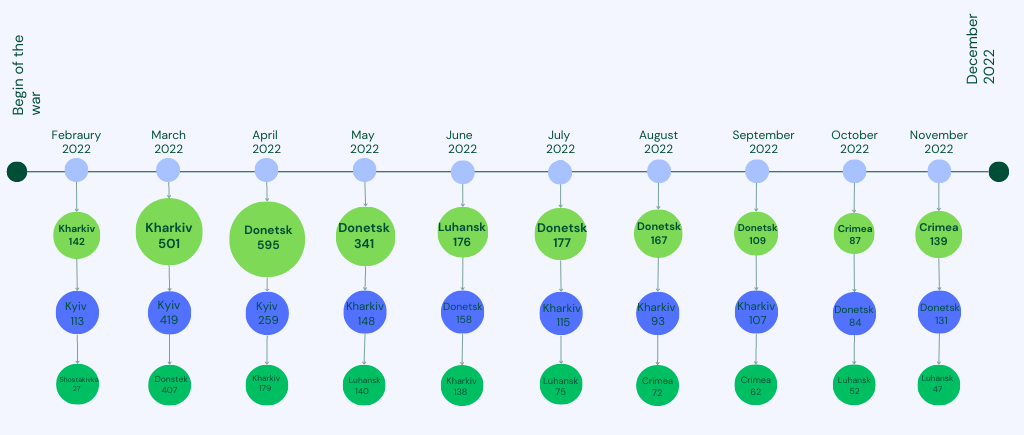}
  \caption{Timeline of top 3 most attacked regions monthly}
  \label{fig:Region-attacks}
\end{figure}

Next, we highlight the three most attacked regions each month. By creating a timeline shown in Figure \ref{fig:Region-attacks} starting in February 2022 until December 2022, these results provide insights into resilience needs as the findings inform policymakers, researchers, and humanitarian organizations about areas requiring targeted support and intervention. It may contribute to the understanding of resilience in Ukraine and guide further research and policy-making.

The SPARQL query \ref{lst:sparql4} 
aims to retrieve distinct address regions. The query applies date filtering, counts the occurrences of each address region, groups the results, and presents the top three regions with the highest occurrence count. The analysis provides insights into the prevalence and distribution of address regions within the dataset, contributing to a better understanding of the geographic aspects of the events being studied.

\begin{lstlisting}[captionpos=b, caption=SPARQL query for getting the most attacked region during certain period of time, label=lst:sparql4,
   basicstyle=\small,frame=single]
PREFIX xsd: <http://www.w3.org/2001/XMLSchema#>
PREFIX rdf: <http://www.w3.org/1999/02/22-rdf-syntax-ns#>
PREFIX rdfs: <http://www.w3.org/2000/01/rdf-schema#>
PREFIX l4r_o: <https://linked4resilience.eu/ontology/>
PREFIX purl: <http://purl.org/dc/terms/>

SELECT distinct ?obj (datatype(?obj) as ?dt)  (count(?sub) as ?numOccurences) WHERE {
  ?sub l4r_o:addressRegion ?obj .
  ?sub purl:date ?date .
  filter(?startDate <= ?date)
  filter(?date < ?endDate)
} group by ?obj  order by desc(?numOccurences)
LIMIT 3
\end{lstlisting}
% how many cities
% how many provinces. 
% which city got the most damage

\subsection{Use Case 5: Ratio of children’s death to monthly attacks }
\label{section:usecase5}
\begin{figure}[ht]
  \centering
  \includegraphics[width=0.85\linewidth]{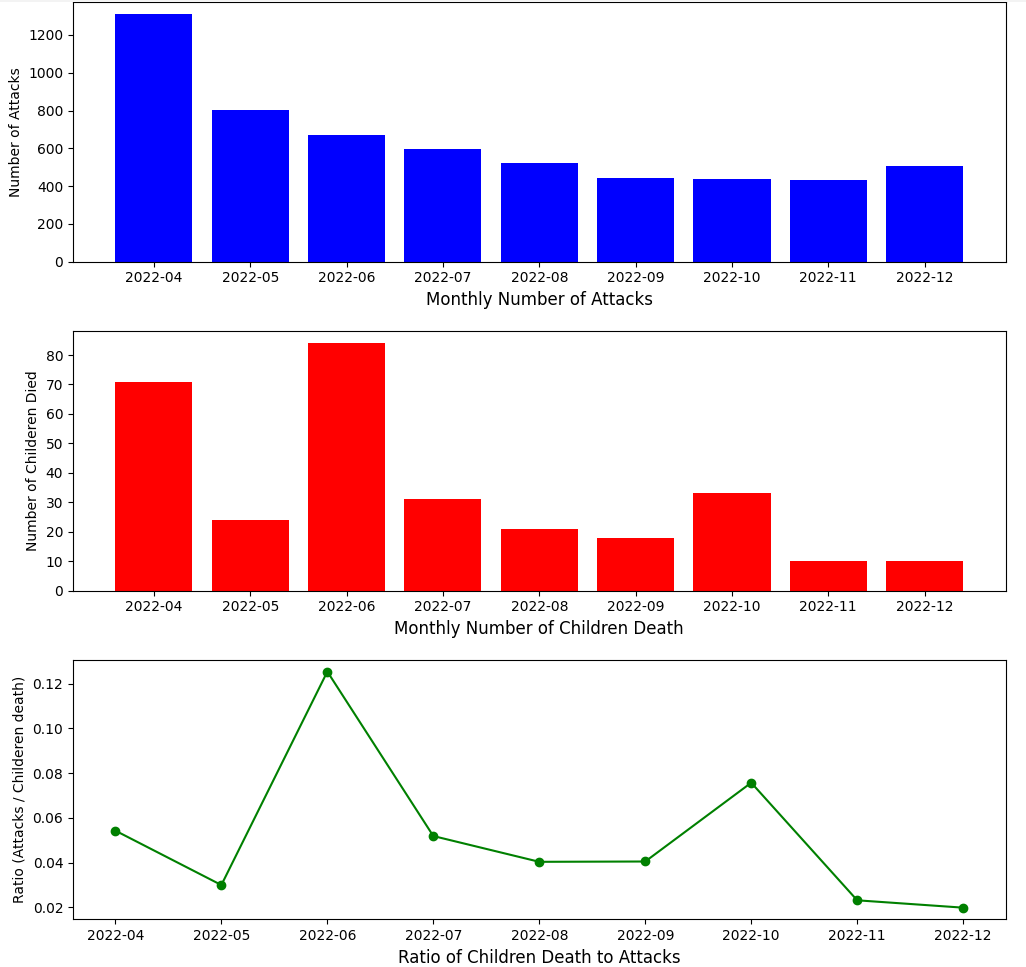}
  \caption{A proof-of-concept study on the ratio of children's death to monthly attacks}
  \label{fig:Monthly_childerent_attack_ratio}
\end{figure}

In addition to the results derived from the output in our integrated data, we illustrate with a proof-of-concept use case how other humanitarian open data such as Uadata \cite{uadata} could be used to conduct further analysis of other aspects in Ukraine. Figure \ref{fig:Monthly_childerent_attack_ratio} presents a monthly record of attacks and children's death between April 2022 and December 2022. The first subplot reveals the monthly number of attacks, shedding light on the intensity of the situation. The second subplot highlights the tragic impact, displaying the monthly count of children who lost their lives. The third subplot captures the severity by depicting the ratio of children's deaths to attacks, offering insights into the relative impact on children. If enriched with complete data, such graphs could serve as a valuable tool for analyzing trends and understanding the gravity of the situation that could provide a reference for humanitarian organizations. However, the data is not complete nor is it easy to examine the reliability of this data. Therefore, this use case merely serves to demonstrate the use. The plots should not be employed in real-world policy decisions.

\subsection{Use Case 6: Identifying location without nearby shelters}
\label{section:usecase6}

\begin{figure}
     \centering
     % \begin{subfigure}[b]{0.45\textwidth}
         \centering
         \includegraphics[width=0.8\textwidth]{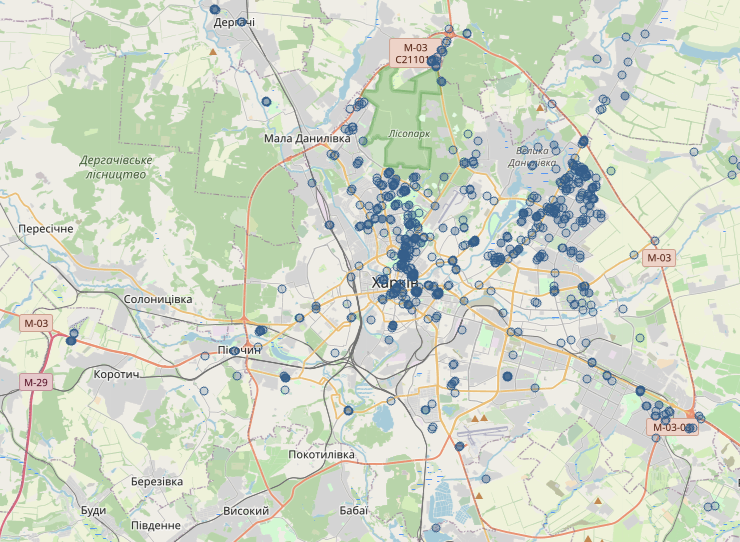}
         \caption{A representation of the damaging events in the city of Kharkiv using YASGUI}
         \label{fig:attack}

     \end{figure}

\begin{figure}[]{}
         \centering
         \includegraphics[width=0.8\textwidth]{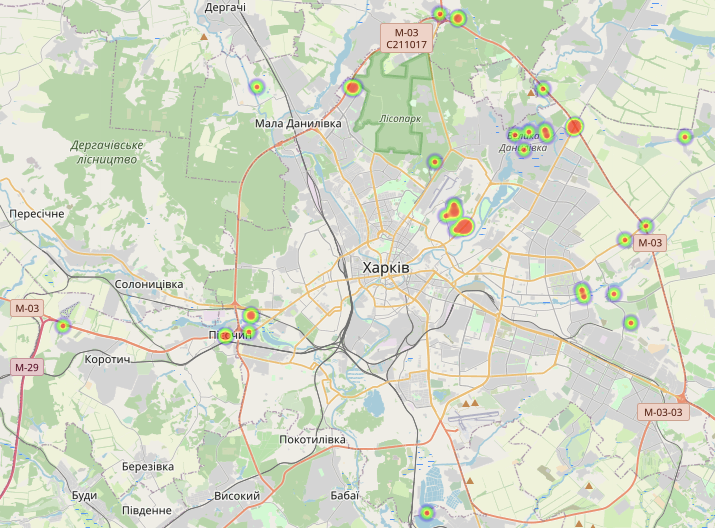}
         \caption{A heatmap regarding the location of attacks in Kharkiv without any shelter within 1km}
         \label{fig:heat}
     % \end{subfigure}
     
        % \caption{Attacks and locations without easy access to shelters}
        \label{fig:three graphs}
\end{figure}

As a demonstration\footnote{A video demo is included in the supplementary material for the use cases \url{https://youtu.be/E_fr1KzfsVs}.} of the use of our integrated dataset, Figure \ref{fig:attack} presents the result of a SPARQL query that retrieves events in Kharkiv in the integrated datasets. 
We retrieved shelter data in the city of Kharkiv \cite{shelters}, and measured the distance of events in Kharkiv to the nearest shelter. Figure \ref{fig:heat} is a heatmap that shows the location of damaging events where there is no shelter within 1km distance. Thus, we suggest that shelters could be built to cover these areas.
\vspace{-2mm}

     % \hfill

% \section{Use Case 7: Updating damaging events using the pipeline}
% \label{section:usecase7}

% {\color{red}We ..... The final integrated dataset consists of X events in Y provinces. We keep the multilingual setting. ...}

\section{Discussion and Conclusion}
\label{section:discussions-conclusion}

This paper presents how existing datasets about damage reporting in Ukraine can be converted to linked data. An algorithm was designed for the automatic detection of identical events, which was used to integrate events from EoR and CH. Our semi-automatic approach reduces ambiguity and enables the enrichment of events with information from other linked open data sources. Finally, we demonstrate with use cases how the resulting integrated dataset can be used for various resilience purposes.

Some invalid links to social media content were detected, including broken links, missing links, and links to content that requires access permission. Our examination shows that 1.1\% and 11.6\% are such invalid links in EOR and CH, respectively. The presence of invalid links in the datasets indicates that the information gathered from social media platforms may not be reliable or complete. This problem is more significant for CH. This limit has been addressed and discussed in some previous research on these of social media data for resilience purposes \cite{limits-crisis-data}. Further assessment and validation are required for the source and resulting datasets.

% This limit has been studied with data on Hurricane Sandy and the Haiti Earthquake \cite{limits-crisis-data}. The study showed that relying solely on social media and phone data to understand people's responses to crisis events has several limitations and may not fully capture the experiences and needs of affected populations. 

% TODO: report how many links that can't be resolved. or redirected, mistaken links (not reporting the same event), etc. 

% volunteer's work have checked them manual validation - The quality could be improved. 
% they can use controlled vocabulary, or ontology to annotate their data. So we don't suffer from spelling errors, confusion due to bilingual entries, etc. 

 While CH is mostly about damaging events, EoR consists of some military events. A beneficial next step is to remove those military events so the content of this dataset is more focused. The resulting dataset could be further enriched with information about the type of buildings, schools, etc. Moreover, the labels such as that of cities, and provinces could be enriched with multilingual information. 
Furthermore, the resulting datasets from this paper project have potential for social impact, by providing valuable insights into the damage caused by the conflict and informing resilience projects. Our data's limitation lies in its inability to precisely capture the conflict's reality. To prevent misuse, measures such as data anonymization, ethical guidelines, and data governance should be implemented before publishing future results openly. In this project, the converted datasets of CH and EoR are available but the resulting integrated data is set to private and only available upon request to avoid the use for unintended proposes.

% Moreover, the analysis based on the use case in Section \ref{section:usecase5} could be inspected in the future by comparing it to records provided by the UN \cite{un_news_ukraine} which brings attention to the tragic reality of children's deaths. It serves as a stark reminder of the profound loss and human tragedy experienced by children in war zones. The data raise an urgent need for concerted efforts to address the protection of the lives of vulnerable children. This prompts critical discussions on policy, intervention, and resource allocation to effectively highlight and prevent child mortality, fostering a global commitment to safeguarding the well-being and future of our children. Additionally, our approach can be extended to include additional datasets, such as ACLED \cite{acled} and WikiEvents \cite{wikievents} to construct a more inclusive and accurate estimation of resilience needs.  And SHACL could be used for further validation. 

Finally, our approach along with the proposed pipeline and use cases is not limited to the specific datasets and context of Ukraine, as it can be adapted for other types of datasets related to geo-annotated events or the resilience of different countries. This adaptability highlights the broader applicability of our methodology in addressing similar challenges worldwide.

% query 3

\subsection*{Acknowledgments}
The authors thank Igor Potapov, Olexandr Konovalov, colleagues, and volunteers for their help.

\printbibliography

@article{victorino2018transforming,
  title={Transforming Open Data to Linked Open Data Using Ontologies for Information Organization in Big Data Environments of the Brazilian Government: the Brazilian Database Government Open Linked Data--DBgoldbr},
  author={Victorino, Marcio and de Holanda, Maristela Terto and Ishikawa, Edison and Oliveira, Edgard Costa and Chhetri, Sammohan},
  journal={Knowledge Organization: KO},
  volume={45},
  number={6},
  pages={443},
  year={2018},
  publisher={Nomos Verlagsgesellschaft mbH und Co KG}
}

@article{limits-crisis-data,
  title={The limits of crisis data: analytical and ethical challenges of using social and mobile data to understand disasters},
  author={Crawford, Kate and Finn, Megan},
  journal={GeoJournal},
  volume={80},
  pages={491--502},
  year={2015},
  publisher={Springer}
}

@inproceedings{wikievents,
  author       = {Vasilis Kopsachilis and
                  Nikos Vachtsavanis and
                  Michail Vaitis},
 
  title        = {WikiEvents - A Novel Resource for NLP Downstream Tasks},
  booktitle    = {Proceedings of the 5th International Workshop on Semantic Methods for Events and Stories, SEMMES, 2023, Hersonissos, Greece, May 29th, 2023},
  series       = {{CEUR} Workshop Proceedings},
  year         = {2023}
}

@article{simpleEventOntology,
  title={Design and use of the Simple Event Model (SEM)},
  author={Van Hage, Willem Robert and Malais{\'e}, V{\'e}ronique and Segers, Roxane and Hollink, Laura and Schreiber, Guus},
  journal={Journal of Web Semantics},
  volume={9},
  number={2},
  pages={128--136},
  year={2011},
  publisher={Elsevier}
}

@article {Chumachenkoo796,
	author = {Chumachenko, Dmytro and Chumachenko, Tetyana},
	title = {Ukraine war: The humanitarian crisis in Kharkiv},
	volume = {376},
	elocation-id = {o796},
	year = {2022},
	doi = {10.1136/bmj.o796

},
	publisher = {BMJ Publishing Group Ltd},
	URL = {https://www.bmj.com/content/376/bmj.o796},
	eprint = {https://www.bmj.com/content/376/bmj.o796.full.pdf},
	journal = {BMJ}
}

@article{acled,
  title={Introducing ACLED: An armed conflict location and event dataset},
  author={Raleigh, Clionadh and Linke, rew and Hegre, H{\aa}vard and Karlsen, Joakim},
  journal={Journal of peace research},
  volume={47},
  number={5},
  pages={651--660},
  year={2010},
  publisher={Sage Publications Sage UK: London, England}
}

@article{kraft2022ethical,
  title={The Ethical Risks of Analyzing Crisis Events on Social Media with Machine Learning},
  author={Kraft, Angelie and Usbeck, Ricardo},
  journal={arXiv preprint arXiv:2210.03352},
  year={2022}
}

@misc{cir,
  author = "{(CIR)}",
  title = "{CENTRE FOR INFORMATION RESILIENCE}",
  year = "{2020}",
  howpublished = "{https://www.info-res.org/}",
  note = "{Accessed on 30-4-2023}"
}

@article{smith2022role,
  author = {Smith, John},
  title = {The Role of Multilingualism in Promoting Resilience and Knowledge Sharing},
  journal = {Journal of Global Resilience},
  volume = {15},
  number = {3},
  year = {2022},
  pages = {123--145}
}

@misc{uadata,
  author = "{uadata.net}",
  title = "{uadata}",
  year = "{Year}",
  howpublished = "{https://github.com/uadata/uadata/blob/main/zlochiny-rf.csv}",
  url = "{https://github.com/uadata/uadata/blob/main/zlochiny-rf.csv}",
  note = "{Accessed on 11 July 2023}"
}

@misc{leedrake5,
  author = "{leedrake5}",
  title = "{leedrake5-Russia-Ukraine}",
  year = "{2022}",
  howpublished = "{GitHub Repository}",
  url = "{https://github.com/leedrake5/Russia-Ukraine}",
  note = "{Accessed on 11 July 2023}"
}

@misc{standwithukraine_2023, 
title={StandWithUkraine}, 
url={https://github.com/vshymanskyy/StandWithUkraine}, 
journal={GitHub}, 
author={vshymanskyy}, 
year={2023}, month={Jun} }

@misc{veelenga_2023, title={GitHub - veelenga/rblist: Russia Ban List}, url={https://github.com/veelenga/rblist}, journal={GitHub}, author={veelenga}, year={2023} }

@misc{shelters, title={shelter-kyiv regional military administration}, url={https://koda.gov.ua/en/the-public/open-data/shelter/}, 
howpublished = "{https://koda.gov.ua/en/the-public/open-data/shelter/}",
journal={Kyiv Regional Military Administration}, year={2022}, month={Jan},note = "{Accessed on 15 July 2023}" }

@inproceedings{sigspatial,
  title={Converting and Enriching Geo-annotated Event Data: Integrating Information for Ukraine Resilience},
  author={Attar, Manar and   Wang, Shuai and  Siebes, Ronald and  Kultorp, Eirik},
  booktitle={ACM International Conference on Advances in Geographic Information Systems (ACM SIGSPATIAL)},
doi={https://doi.org/10.1145/3589132.3625580},
  year={2023},
note = {Updated datasets, use cases and demos are available at \url{https://linked4resilience.eu/} (accessed 2023-10-04).}
}

\end{document}